\documentclass{l4dc2025}

\title[Implicit Game-Theoretic MPC]{Learning Two-agent Motion Planning Strategies from Generalized Nash Equilibrium for Model Predictive Control
}
\usepackage{multirow}
\usepackage{times}
\usepackage{algorithm}
\usepackage[noend]{algpseudocode}
\usepackage{amsmath, amssymb} 

\author{%
 \Name{Hansung Kim}\textsuperscript{1} \Email{hansung@berkeley.edu}\\
 \Name{Edward L. Zhu}\textsuperscript{2} \Email{edward.zhu@plus.ai}\\
  \Name{Chang Seok Lim}\textsuperscript{1} \Email{cshigh22@berkeley.edu}\\
  \Name{Francesco Borrelli}\textsuperscript{1} \Email{fborrelli@berkeley.edu}\\
 \addr \textsuperscript{1}University of California, Berkeley \\
 \addr \textsuperscript{2}PlusAI, Inc
}

\begin{document}

\maketitle

\begin{abstract}%
We introduce an Implicit Game-Theoretic MPC (IGT-MPC), a decentralized algorithm for two-agent motion planning that uses a learned value function that predicts the game-theoretic interaction outcomes as the terminal cost-to-go function in a model predictive control (MPC) framework, guiding agents to implicitly account for interactions with other agents and maximize their reward. This approach applies to competitive and cooperative multi-agent motion planning problems which we formulate as constrained dynamic games. Given a constrained dynamic game, we randomly sample initial conditions and solve for the generalized Nash equilibrium (GNE) to generate a dataset of GNE solutions, computing the reward outcome of each game-theoretic interaction from the GNE. The data is used to train a simple neural network to predict the reward outcome, which we use as the terminal cost-to-go function in an MPC scheme. We showcase emerging competitive and coordinated behaviors using IGT-MPC in scenarios such as two-vehicle head-to-head racing and un-signalized intersection navigation. IGT-MPC offers a novel method integrating machine learning and game-theoretic reasoning into model-based decentralized multi-agent motion planning.
\end{abstract}

\section{Introduction}
\label{sec:introduction}
\begin{figure}
    \centering
    \includegraphics[width=0.5\columnwidth]{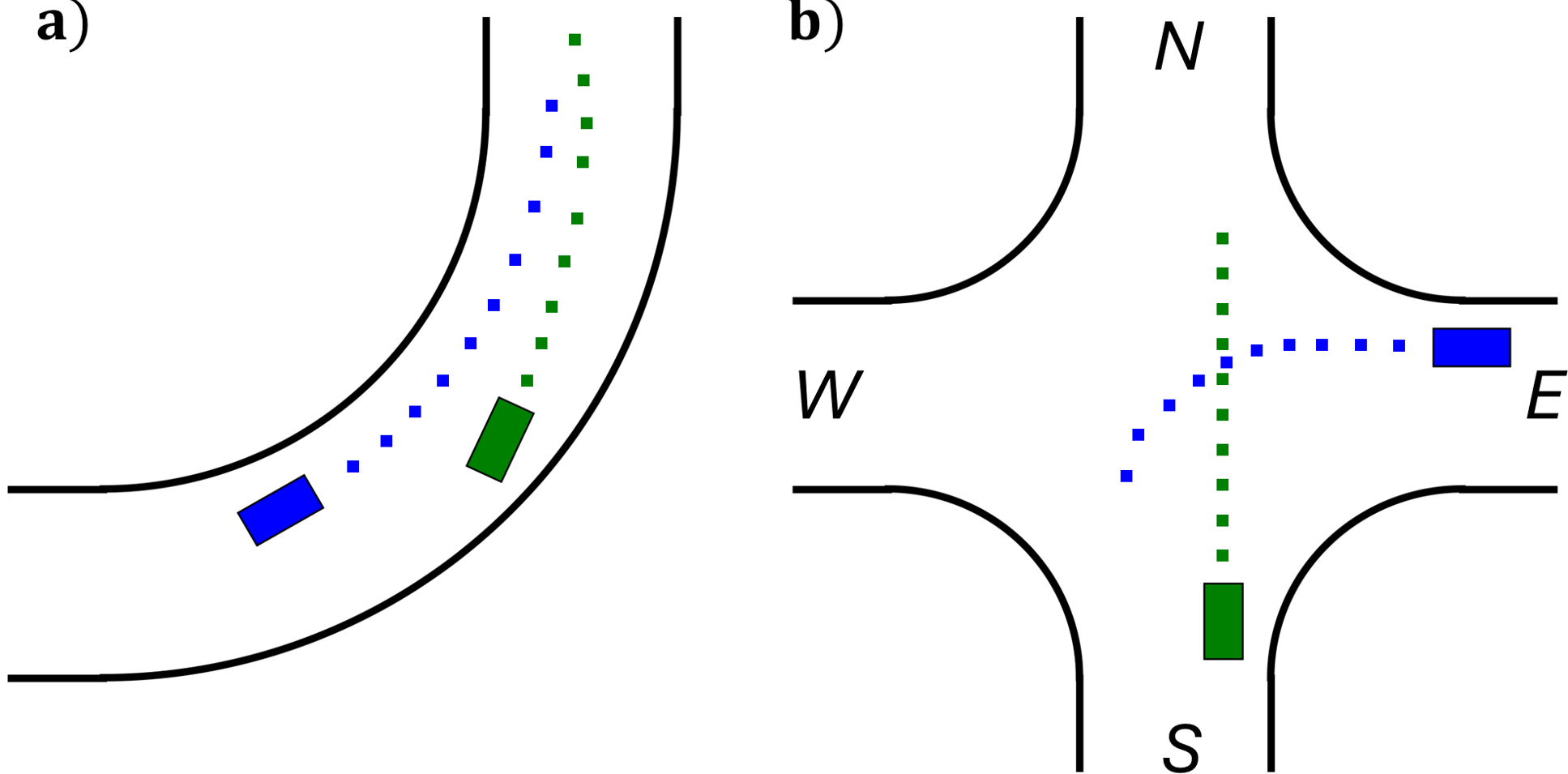}
    \caption{Two-agent interaction scenarios: a) competitive head-to-head racing, and b) cooperative un-signalized two-way intersection navigation, with colored squares showing vehicles' planned trajectories.}
    \label{fig:scenarios}
    \vspace{-0.7 cm}
\end{figure}

 Multi-agent motion planning is a fundamental challenge for autonomous systems navigating complex and dynamic environments, especially when an ego agent must adapt in real-time to the actions of other agents.  Multi-agent settings are generally categorized as competitive, cooperative, or non-cooperative. In competitive settings, agents pursue conflicting goals, often seeking outcomes that favor themselves at the expense of their opponents. An example, as shown in Fig. \ref{fig:scenarios}a, is head-to-head car racing, where competing agents attempt to finish a race ahead of their opponent. In cooperative settings, agents collaborate to achieve shared objectives, such as navigating an un-signalized intersection with connected autonomous vehicles (CAVs), which must coordinate to move safely and efficiently without causing delays or collisions, as shown in Fig. \ref{fig:scenarios}b.
 Lastly, in non-cooperative settings, agents act independently to maximize their own interests without collaboration. While our approach applies to all three multi-agent environment settings, we will demonstrate our approach on the two examples illustrated in Fig. \ref{fig:scenarios}. The main source of challenge in multi-agent motion planning is the complex interaction between agents in a shared environment. Robotaxi services, now operational in select U.S. cities, showcase advanced autonomous navigation but occasionally face the ``freezing robot" problem—vehicles freezing to avoid collisions—highlighting challenges in real-world interactions between robots \citep{robotaxi_article, waymo_standoff}. Decentralized non-convex algorithms with coupled collision avoidance constraints, as demonstrated in \cite{jankovic}, are prone to instability, while purely decentralized strategies without interaction modeling often lead to gridlock. In contrast, interaction-aware decentralized approaches with best-response computations by accounting for other agents' actions effectively avoid these issues.

 A common method for interaction-aware motion planning leverages game-theoretic principles to explicitly model multi-agent interactions, accounting for how each agent's actions influence others. As in \cite{zhu2024,Wang2018GameTM,potential-ilqr}, the multi-agent motion planning problem is posed as an equilibrium finding problem of a constrained dynamic game, for which the generalized Nash equilibrium (GNE)—a set of strategies where no agent has an incentive to unilaterally deviate from its GNE strategy—is a common solution concept. In \cite{zhu2024}, a novel numerical solver for computing GNE for open-loop dynamic games with state and input constraints is proposed and the solver is demonstrated in autonomous racing examples. However, this method suffers from computational efficiency which limits its use for real-time planning. Similarly, \cite{Wang2018GameTM} proposed a game-theoretic formulation for multi-robot racing that approximates the GNE using an iterative best-response method but also faces scalability challenges. The Potential-iLQR approach \citep{potential-ilqr} formulates multi-agent interactions as a potential dynamic game \citep{potential_games} and uses distributed optimization to approximate the GNE in a scalable manner. However, the potential game formulation restricts the application of this method to cooperative interactive scenarios and cannot be used to describe games with zero-sum components in their agent objective functions. Furthermore, the game-theoretic motion planning approaches presented in these works require precise knowledge of other agents' dynamics, costs, and decision-making processes \citep{zhu2024thesis}, often requiring assumptions about other agents' behavior that may not hold true in real-world scenarios. This limitation motivates our approach to design a feedback policy that uses game-theoretic reasoning as a guiding strategy while adapting to the observed behaviors of other agents by using a learned value function to integrate game-theoretic reasoning into a feedback MPC policy via the terminal cost function.

In this paper, we present an algorithm for learning two-agent motion planning strategies from a dataset of game-theoretic interactions and embedding it in a real-time model predictive control (MPC) scheme to guide the solution of short-horizon optimal control problems. Our approach comprises three main steps:
\begin{enumerate}
    \item \textbf{Game-theoretic Interaction Dataset Generation:} Generate the dataset offline by randomly sampling joint initial conditions and solving constrained dynamic games over agents and storing the GNE solutions.
    \item \textbf{Reward Outcome Learning:} Train a neural network value function to predict the reward outcomes associated with these interactions. The reward outcome captures the objective of the multi-agent task in a low-dimension. For example, in racing scenarios, the reward reflects progress advantage along a racetrack relative to the opponent; in cooperative navigation, it represents collective progress toward the agents' target positions.
    \item \textbf{Implicit Game-Theoretic MPC (IGT-MPC):} Use the learned value function as the terminal cost-to-go function in an MPC policy for online control.
\end{enumerate}
By integrating a learned value function that encodes the strategic outcomes arising from game-theoretic motion planning, IGT-MPC enables implicit interaction-aware motion planning without the need for explicit interaction models or computationally intensive game-theoretic equilibrium finding \citep{zhu2024,Wang2018GameTM}. 
\begin{remark}
    The current IGT-MPC implementation focuses on two-agent motion planning problems, leaving open the challenges of scaling to larger numbers of agents and generalizing to unseen environments such as varying race track layouts or intersection geometries. While this initial effort does not address these challenges, we acknowledge their importance for enabling practical real-world applications. 
    The primary aim of this work is not to provide an immediate solution for autonomous racing or driving but to demonstrate that game-theoretic strategic outcomes can be effectively encoded in neural networks which can then be integrated into numerical optimization schemes for model-based optimal control, paving the way for research in integration of game-theoretic reasoning, machine learning, and model-based optimal control.
\end{remark}

\vspace{-0.3 cm}
\section{Problem Formulation}\label{sec:prblm_f}
Consider a two-agent autonomous vehicle system where the finite-horizon, discrete-time dynamics are described by 
\small\begin{align}\label{eq:joint_dyn}
    z_{t+1}=f(z_t,u_t),
\end{align}\normalsize
where $z^i_t\in\mathcal{Z}^{i}, u^i_t\in\mathcal{U}^{i}$ are the state and input of agent $i$ at time $t$ and 
\begin{align}\label{eq:XandU}
    z_t:=[z^1_t,z^2_t]^\top \in \mathcal{Z}^1 \times\mathcal{Z}^2=\mathcal{Z}\subseteq\mathbb{R}^{2n_z},\\
    u_t:=[u^1_t, u^2_t]^\top \in \mathcal{U}^1 \times\mathcal{U}^2=\mathcal{U}\subseteq\mathbb{R}^{2n_u},
\end{align}
are the concatenated states and inputs of all agents.

Agent $i$ is subject to state and input constraints defined by
$\mathcal{Z}^i=\{z\in \mathbb{R}^{n_z} \ | \ \underline{z}^i \leq z \leq \overline{z^i}\}$ and 
$\mathcal{U}^i=\{u\in \mathbb{R}^{n_u} \ | \ \underline{u}^i\leq u \leq \overline{u^i}\}$, where $\underline{x}$ and $\overline{x}$ are lower and upper bounds on the respective variables which can be derived from the vehicle's actuation limitations, traffic law, or road boundaries. Also, agent $i$ is subject to actuation rate limits denoted by 
$\Delta\mathcal{U}^i=\{\Delta u\in \mathbb{R}^{n_u} \ | \ \underline{\Delta u}^i\leq \Delta u \leq \overline{\Delta u^i}\}$, where $\Delta u_k:=[u^1_k-u^1_{k-1}, u^2_k-u^2_{k-1}]^\top \in \Delta\mathcal{U}^1 \times \Delta \mathcal{U}^2 = \Delta \mathcal{U} \subseteq \mathbb{R}^{2n_u}$.

Each agent $i$ minimizes its own cost function, which is comprised of stage cost $\ell_k^i(z_k,u^i_k)$ and terminal cost $\ell_{N_G}^i(z_{N_G})$ over a finite horizon length of $N_G$: 
\begin{subequations}\label{eq:cost_i}
\begin{align}
    J^i(\mathbf{z},\mathbf{u^i})&=\sum_{k=0}^{N_G-1} \ell_k^i(z_k,u^i_k) + \ell_{N_G}^i(z_{N_G}) \label{eq:cost_z_u} \\
    &= J^i(\mathbf{u}^i,\mathbf{u}^{\neg i},z_0), \label{eq:cost_u_i}
\end{align}
\end{subequations}
where the $\mathbf{u^i}=[u^i_0\dots,u^i_{N_G-1}]$ denote the input sequences of agent $i$, and $\mathbf{z}=[z_0,\dots,z_{N_G}]$ is the sequence of joint states over the prediction horizon $N_G$. In this work, we use the notation $z^{\neg i}_k$ and $u^{\neg i}_k$ to denote the collection of states and inputs for all but the $i$-th agent. We derive \eqref{eq:cost_u_i} by recursively substituting the dynamics \eqref{eq:joint_dyn} into the cost function \eqref{eq:cost_z_u}, which inherently depends on the open-loop input sequences of all agents. The cost is a function of the joint state and input sequence of the $i$-th agent, capturing dependence on the behavior of other agents. The agents are also subject to additional $n_c$ constraints, $C(\mathbf{u}^1,\mathbf{u}^2,z_0)\leq 0$, which describe the coupling between agents, with its dependence on joint dynamics remaining implicit.
\section{Learning Game-theoretic Motion Planning Strategy from GNE}
In this section, we describe our approach to learning game-theoretic motion planning strategies from GNE and how to embed it in an MPC policy for motion planning in competitive and cooperative two-autonomous vehicle systems: 1) two-vehicle head-to-head racing on an L-shaped track. 2) two-vehicle navigating through an un-signalized four-way intersection. 

\subsection{Game-theoretic Interaction Data Generation}
First, we define the constrained dynamic game as the tuple:
\begin{align}\label{eq:game_tuple}
    \Gamma = (N_G,\mathcal{Z},\mathcal{U},\Delta\mathcal{U},f,\{J^i\}_{i=1}^2,C).
\end{align}
For such a game, a GNE \citep{Facchinei2010} is attained when all agents have no incentive to unilaterally change their strategies along feasible directions of the solution space, which means $\mathbf{u}^\star=[\mathbf{u}^{1,\star},\mathbf{u}^{2,\star}]$ minimizes \eqref{eq:cost_i} for both agent 1 and 2. There exists a vast literature on formulating constrained dynamic games for different robotic multi-agent motion planning problems and solving for GNE, such as ALGAMES, iterative Linear Quadratic Regulator (iLQR), Potential-iLQR, and DG-SQP \citep{algames,iLQR,potential-ilqr,zhu2024}. Without loss of generality, we use the DG-SQP solver~\citep{zhu2024} to compute a GNE for constrained dynamic games in this work. 


We randomly sample joint initial conditions $z_0$ while other components of the game $\Gamma$ are fixed and solve for the GNE from that initial condition. Then, the GNE is used to compute the reward outcome of game-theoretic interactions. The reward $R^i\in\mathbb{R}$ quantifies the success in achieving the primary objective of the multi-agent interactive motion planning problem from the perspective of Agent $i$. While $J^i$ in \eqref{eq:cost_i} describes Agent $i$'s cost function and includes stage costs $\ell_k^i$, $R^i$ is derived from the terminal costs $\ell_{N_G}^i,\; i=1,2$, and thus directly reflects the game outcome.

\subsubsection{Two-vehicle head-to-head racing}
Here the objective is to stay ahead of the opponent. We use the reward
\begin{align} \label{eq:racing_reward}
R^i=s^i_{N_G}-s^{\neg i}_{N_G}, \quad i=1,2,
\end{align}
that represents the advantage of vehicle $i$ over vehicle $\neg i$, where $s^i_{N_G}$ and $s^{\neg i}_{N_G}, $ are the longitudinal progress of Agent $i$ and Agent $\neg i$ on the race track at time step $N_G$, respectively. When $i=1$, we are considering Agent 1 as the ego agent, which means we are predicting the progress advantage that Agent 1 will have over Agent 2. We also construct the reward from the perspective of Agent 2 as the ego agent when $i=2$. Thus, from a single GNE solution, we extract two data points from the perspective of each agent. Each reward target $R$ is associated with an input feature vector which is the joint initial condition augmented with additional information about the interaction. We note that the input vectors are also problem-dependent, and in the two-vehicle head-to-head racing example, we define 
\begin{align}\label{eq:race_feature}
 \Tilde{x}^i_{0} = [z^{\neg i}_0,z^i_0-z^{\neg i}_0]^\top \in \mathbb{R}^{n_x} \quad i=1,2.
\end{align} 
Here, $z^i_0 =[v^i_{0},e^i_{\psi,0},s^i_0,e^i_{y,0}]$, where $v^i_0$ denotes the initial longitudinal velocity, $e^i_{\psi,0}$ is the heading angle error, $s^i_0$ represents the progress along the race line, and $e^i_{y,0}$ is the lateral error of agent $i$. 

\subsubsection{Collaborative navigation through an un-signalized intersection}
The main objective of the agents is to collaborate with other agents to ensure all agents pass the intersection safely and promptly. In this example, we propose the reward
\begin{align} \label{eq:int_reward}
R^i=\sum\nolimits_{i=1}^{2} s^i_{N_{G}} \quad i =1,2,    
\end{align}
which represents the collective progress along corresponding routes at the end of the game horizon $N_G$. The associated input feature to $R^i$ is  
\begin{align}\label{eq:int_feature}
 \Tilde{x}^i_{0} = [\Tilde{z}^{\neg i}_0,\Tilde{z}^i_0-\Tilde{z}^{\neg i}_0]^\top \in \mathbb{R}^{n_x} \quad i=1,2,
\end{align} 
where $\Tilde{z}^i_0 =[s^i_{0},v^i_{0},sc_i]$. Assuming all agents do not deviate from the centerline of the road, we only include the longitudinal states and scenario encoding $sc_i$, which is a categorical variable, of the two-vehicle intersection to distinguish what type of interaction is taking place in the perspective of agent $i$. Without it, data points from different interactions (i.e. left turn \& left turn vs. left turn \& straight) become indistinguishable. For more details about scenario encoding and its implementation, please refer to Sec. \ref{sec:int_implemnetation}. Similarly, we construct the features and reward targets from the perspective of both agents. Note that the reward target for each agent is the same, which reflects the cooperative setting where all agents share the reward. The computed reward $R=\{R^i\}_{i=1}^2$ and feature vector $\Tilde{x}_0=\{\Tilde{x}^i_0\}_{i=1}^2$ are added to the dataset as follows 
\small{\begin{align}\label{eq:dataset}
\begin{split}
    X &= X\cup \{\Tilde{x}_0\} \\
    Y &= Y \cup \{R\}, 
    \end{split}
\end{align}}\normalsize

\begin{remark}
    Capturing interaction between agents in the generated dataset is key for a successful implementation of the proposed method. In practice, the distribution of training data often differs from the distribution seen during IGT-MPC deployment, a phenomenon known as distribution shift. As covering the entire feature space by exhaustively sampling data points is intractable, a more data-efficient approach leverages strategic initial condition sampling based on any prior knowledge of the deployment data distribution. For example, in a head-to-head racing scenario, when two agents are far enough apart, then there isn't any significant coupling between their decision-making. Thus, sampling from the set of initial conditions where an interaction between two agents is possible will be sufficient in capturing the interactive behavior. 
\end{remark}

\begin{remark}
When sampling initial conditions, the game solver may occasionally fail to compute the GNE due to infeasibility from collision avoidance constraints under the joint dynamics. Storing these infeasible points in the dataset and associating them with low reward outcomes is crucial, as they represent undesirable states leading to collisions that the model must learn to avoid.
\end{remark}

\subsection{Learning the Reward Outcome}
After generating the dataset $\mathcal{D}=(X,Y)$, we obtain the game-theoretic value function $V_{GT}:\mathbb{R}^{n_x} \mapsto \mathbb{R}$ that estimates the reward outcome $R^i$ of the game-theoretic interactions between agents at time step $N_G$ from the input features $\Tilde{x}^i_0$. We use the following simple neural network architecture with $tanh$ activation function:
\begin{equation}
   { y_l = tanh(W_l y_{l-1} + b_l), \quad l = 1, ... , L},
\end{equation}
where $W_l, b_l$ are the parameters of the $l$-th layer of the network with hidden state dimension $h$. Additionally, $y_0=\Tilde{x}^i$ and $y_L=\hat{R}^i$, and $tanh$ is a non-linear, differentiable activation function ideal for expressing negative-valued outputs.

The regression model is trained on $\mathcal{D}$ using the mean-squared error loss function defined as $\mathcal{L}=\frac{1}{n_b} \sum_{k=1}^{n_b}\sum_{i=1}^{2}(R^i_k - \hat{R^i}_k)^2$ where $n_b$ is the training batch size. We employ the Adam optimizer \citep{AdamAW} with early stopping to prevent overfitting. Also, the features and target values in $\mathcal{D}$ are normalized during training. From this point forward, the trained neural network value function on a game-theoretic interaction dataset is referred to as $V_{GT}$. Notably, this relatively simple neural network architecture effectively captures game-theoretic strategies in the GNE for two-agent motion planning in both competitive and cooperative settings, as demonstrated in Sec. \ref{sec:examples}. We note however, that our current approach focuses on two-agent systems and is limited in generalizability to unseen environments (e.g., different racetrack or intersection geometries). To enhance scalability w.r.t. the number of agents and generalizability, the interactions among all agents and environmental information must be incorporated into the feature vector, which may require more complex neural network architectures—a direction we plan to explore in future work.

\subsection{Implicit Game-theoretic MPC Policy}
After training $V_{GT}$ that predicts the game-theoretic reward outcome $N_G$ steps into the future given the input features, we use this predictor as a terminal cost-to-go function for a real-time MPC policy that implicitly emulates the game-theoretic strategies encoded in GNE. 

\subsubsection{Collision Avoidance}
While methods like optimization-based collision avoidance constraints \citep{obca}, which model agents as polytopes, can be used, we opt to represent obstacles as circles and impose distance constraints, maintaining simplicity without compromising effectiveness. Inter-agent collision avoidance constraints are given as $\mathbb{CA} = \{(p_1,p_2) \ | \ \lVert p_1-p_2\rVert_2 \geq d_{min}\}$, where $p_1$ and $p_2$ are global Cartesian position vectors, and $d_{min}$ is the user-defined parameter for minimum allowable distance between any two agents. We denote $P(z^i,z^j)$ as the projection of the state vector pair to Cartesian position vector space. Two agents $i$ and $j$ are considered to be collision-free if and only if $P(z^i,z^j) \in \mathbb{CA}$.
\vspace{-0.3 cm}
\subsubsection{MPC Formulation}
The MPC policy is a \textit{predict-then-plan} scheme that computes an optimal ego trajectory w.r.t. the open-loop predictions of the environment. We assume the forecasts are provided by a trajectory prediction scheme. In the numerical examples, we generate $\mathbf{\hat{z}}^{\neg i}$ using an oracle that provides the optimal planned trajectories of all agents. For competitive racing scenarios, we corrupt it with Gaussian noise \citep{zhu2024thesis}. 
 
Each agent $i$ computes a state-feedback control $u_t^i=\pi_{\text{MPC}}(z_t^{i},\mathbf{\hat{z}}^{\neg i},V)$ by solving the following finite-horizon optimal control problem (FHOCP) towards tracking a state reference $\{z_{r,k}\}_{k=t}^{t+N}$ given forecasts of other agents $\mathbf{\hat{z}}^{\neg i}=\{\hat{z}_{k|t}^{\neg i}\}_{k=t}^{t+N}$ over the $N$ step MPC prediction horizon,
{\small{
\begin{subequations}\label{opt:mpc_pol}
\begin{align}
 \min_{\boldsymbol{u^i}}&\; \sum\limits_{k=t}^{t+N-1}\! \lVert z^{i}_{k|t}-z_{r,k} \rVert_Q^2 + \lVert u^i_{k|t} \rVert^2_{R_1} + \lVert \Delta u^i_{k|t} \rVert^2_{R_2} + V({\Tilde{x}^i_{t+N}})
 \label{eq:cost}\\
 \text{s.t. }&\quad z^i_{k+1|t}=f^i(z^i_{k|t},u^i_{k|t}), \quad \forall k\in\mathbb{I}_t^{t+N-1}\\
 &\quad z^i_{k|t} \in \mathcal{Z}^i, \quad \forall k\in\mathbb{I}_t^{t+N}\\
 &\quad u^i_{k|t} \in \mathcal{U}^i, \quad \forall k\in\mathbb{I}_t^{t+N-1}\\ 
 &\quad \Delta u_{k|t}^i \in \Delta \mathcal{U}^i, \quad \forall k\in\mathbb{I}_t^{t+N-1}\\
 &\quad P(z^i_{k|t},\hat{z}^j_{k|t})\in \mathbb{CA}, \quad \forall j\in \{\neg i\}\\
 &\quad z^i_{t|t} = z^i(t),
\end{align}
\end{subequations}}}where $z^i(t)$ is the measured state of the $i$-th agent at time $t$, and $z^i_{k|t}$ indicates the state $z^{i}$ at $k$-th step predicted at time $t$. $f^i$ is the discrete-time dynamics of agent $i$, and $\mathbb{I}_{k_1}^{k_2}$ denote the index set $\{k_1,k_1+1,\dots, k_2\}$. $z_r$ is a reference state trajectory and $Q\succeq 0$ and $R_1,R_2 \succ 0$ are weighting matrices for the state reference tracking, input, and input rate costs respectively. The reference state trajectory is the pre-computed raceline or the centerline of a road.

The cost function \eqref{eq:cost} includes a stage cost term $\ell_k^i(\cdot)$ and a terminal cost term $V(\cdot)$. It is important to note that the input to the terminal cost function $\Tilde{x}^i_{t+N}$ includes the joint state $z_{t+N|t}=[z^i_{t+N|t},\hat{z}^{\neg i}_{t+N|t}]$ where $z^i_{t+N|t}$ is a decision variable of the optimization problem. Therefore, if $V=-V_{GT}$, minimizing the learned value function over $z^i_{t+N|t}$ can be interpreted as agent $i$ selecting its terminal state given the forecasts of other agents, $\hat{z}^{\neg i}_{t+N|t}$, to maximize its reward (i.e. maximize advantage over the opponent). The learned value function, $V_{GT}$, is used in closed-loop with the MPC policy as formalized in Algorithm \ref{alg:proposed_alg}. The FHOCP \eqref{opt:mpc_pol} is constructed using CasADi \citep{casadi} where $V_{GT}$ is constructed symbolically using the same architecture and trained weights from the learned model in PyTorch \citep{pytorch}.
\small{
\begin{algorithm}
\caption{\textbf{Implicit Game-Theoretic MPC (IGT-MPC)}}\label{alg:proposed_alg}
\begin{algorithmic}
\Require $V_{GT}$
\State Set the task horizon $T$
\State $t\gets 0$ \\
\While{$t < T$}{
    \State $\mathbf{\hat{z}}^{\neg i} \gets$ Forecasts of other agents
    \State $\Tilde{x}_{t+N} \gets \mathbf{\hat{z}}^{\neg i}$
    \State $V\gets V_{GT}$
    \State $u^i_t \gets \pi_{\mathrm{\text{MPC}}}(z^i(t),\mathbf{\hat{z}}^{\neg i},V)$ \Comment{See Eq.~\eqref{opt:mpc_pol}}
    \State Apply $u_t^i$ to system
    \State $t \gets t+1$}
\end{algorithmic}
\end{algorithm}
}

\section{Numerical Examples} \label{sec:examples}
\label{sec:results}
We demonstrate Algorithm \ref{alg:proposed_alg} in two-vehicle competitive and cooperative multi-agent settings. The code for obtaining the presented results can be found in the \href{https://github.com/MPC-Berkeley/Implicit-Game-Theoretic-MPC}{GitHub repository} \citep{github}. We use $N=10$ and a sampling time of $0.1\;s$ for all MPC policies in our numerical examples. 


\subsection{Two-vehicle Head-to-Head Racing}
In the two-vehicle head-to-head race example, we consider the Frenet-frame dynamic bicycle model given a precomputed race line trajectory for an L-shaped race track \citep{zhu2024}. 
We formulate the autonomous racing problem as a constrained dynamic game and use DG-SQP solver to compute the GNE for such game. For an in-depth explanation of formulating the game and solving for GNE, refer to \cite[Chapter 7]{zhu2024thesis}. We collected approximately 1800 samples with $N_G=25$, and trained $V_{GT}$ with $h=48, L=2$ using PyTorch \citep{pytorch}. 

The MPC policy \eqref{opt:mpc_pol} is additionally subject to performance limiting factors such as friction limits from nonlinear tire models as follows
\begin{align}\label{eq:racing_friction}
    a^f(z^i_{k|t},u^i_{k|t}), a^r(z^i_{k|t},u^i_{k|t}) \leq \mu g,
\end{align} 
where $a_f(\cdot,\cdot)$ and $a_r(\cdot,\cdot)$ denote the front and rear tire lateral acceleration, respectively. $\mu$ is the coefficient of friction of the race track, and $g$ is the gravitational acceleration. 

To study the impact of the $V_{GT}$, we consider two different terminal cost-to-go functions,
  \begin{equation} \label{eq:V_race}
    V=
    \begin{cases}
      -V_{GT}(z^i,\hat{z}^{\neg i}), \\
      V_{MP}(z^i,\hat{z}^{\neg i})=-(s^i - \hat{s}^{\neg i}),\quad i = 1,2,
    \end{cases}
  \end{equation}
 where $V_{MP}$ is a naive maximum progress function based on the opponent's prediction to encourage progress maximization with respect to the opponent but does not have any game-theoretic component.

\begin{figure*}
    \centering
    \includegraphics[width=0.9\textwidth]{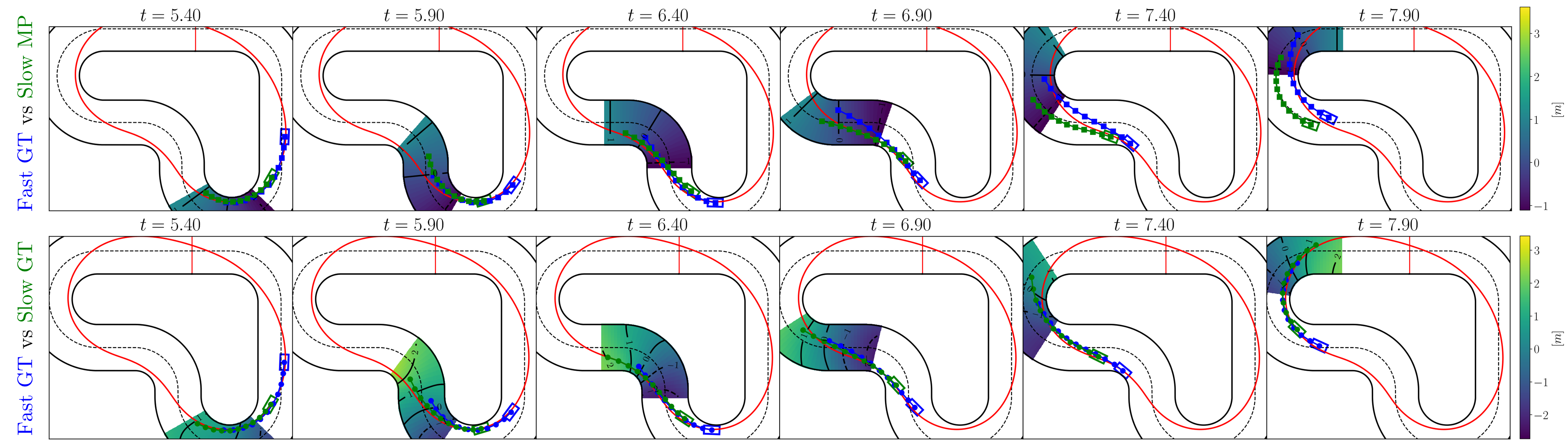}
    \caption{\small{In a simulation experiment, the slower (green) vehicle is unable to defend its position against the faster (blue) vehicle with $V_{MP}$ (Top). The slower vehicle with $V_{GT}$ successfully defends its position against the faster car with $V_{GT}$ (Bottom). The heatmap represents the level curves to visualize the value functions used in this experiment. The red curve is the raceline and the colored squares and circles are the planned trajectories for vehicles with corresponding colors.}}
    \label{fig:racing_time_trace}
\end{figure*}
\subsubsection{Numerical Results}
Fig. \ref{fig:racing_time_trace}
shows snapshots at identical time instances of two simulated races starting from the same initial condition but different configurations to demonstrate a position defense capability of a slower vehicle when guided by $V_{GT}$. Race configurations are categorized by two settings: \textit{fast} or \textit{slow} and \textit{GT} or \textit{MP} value functions. The \textit{GT} or \textit{MP} setting denotes the terminal cost-to-go function used by the MPC policy controlling the vehicle. While both vehicles track pre-computed racelines of the same shape, we force the initially leading green vehicle to be slower than the initially trailing blue vehicle by reducing its velocity profile by 10\%. For details on setting up the race, refer to \cite[Chapter 7]{zhu2024thesis}. The level curves of $V_{MP}$ (Top) and $V_{GT}$ (Bottom) for the green vehicle are projected onto the track in Fig. \ref{fig:racing_time_trace}. While the level curves of $V_{MP}$ are constant over lateral positions, indicating no preference for where on the track the vehicle should end up at the end of the horizon, $V_{GT}$ exhibits more intricate level curves, showing a strategic preference for lateral positions near the pre-computed race line, although the learned value functions had no prior knowledge of it \citep{zhu2024thesis}. By $t=7.90\;s$, the green vehicle guided by $V_{MP}$ is overtaken by the faster opponent, while under $V_{GT}$, it successfully defends its leading position. More simulation results can be found in the \href{https://youtu.be/9jlz95Nor2I}{video} \citep{youtube_vid}.

 \begin{figure*}
    \centering \includegraphics[width=0.9\textwidth]{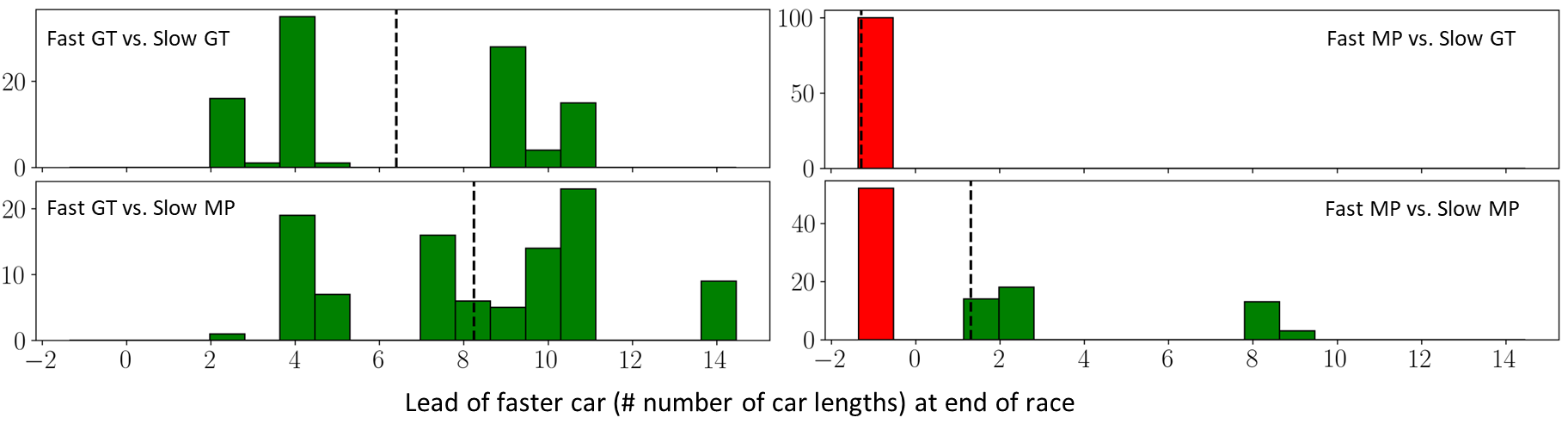}
    \caption{\small{Histograms of the lead (in number of car lengths) of the faster vehicle over the slower vehicle at the end of each simulated race over 100 different initial conditions. Green bars indicate bins where the faster vehicle won the race, while red bars indicate losses. The black dashed lines represent average lead values over all simulation runs. Note that vertical scales vary between histograms.}}
    \label{fig:race_stats}
    \vspace{-1 cm}
\end{figure*}

Now, we report the Monte Carlo simulation results for various race configurations over 100 different initial conditions. The histograms display the lead of the faster car, measured in car lengths, at the end of each race across 100 different initial conditions for various race configurations in Fig. \ref{fig:race_stats}. The black dashed lines indicate the mean values. Comparing the histograms in the top row, the \textit{Fast GT} vehicle consistently wins averaging approximately $6.5$ car lengths of a lead against \textit{Slow GT} vehicle, while the \textit{Fast MP} vehicle trails the \textit{Slow GT} by an average of $1.4$ car lengths. Notably, \textit{Fast MP} vehicle loses $100 \%$ of the race against a slower, but more strategic vehicle guided by $V_{GT}$. In the bottom row, a faster vehicle races against a \textit{Slow MP} vehicle. When guided by $V_{GT}$, the faster vehicle is able to win $100 \%$ of the races while averaging $8.1$ car lengths of a lead. However, with the $V_{MP}$, the win rate is decreased to about $50 \%$. Overall, $V_{GT}$ outperforms the vehicle using $V_{MP}$ in both overtaking and defending, as highlighted in the \textit{Fast GT vs Slow MP} and defending in \textit{Fast MP vs Slow GT} cases.
\subsection{Two-vehicle Intersection Navigation}
In the two-vehicle un-signalized intersection example, we consider the Frenet-frame kinematic bicycle model \citep{KIM2023} augmented with global Cartesian coordinates $(x,y)$ for imposing the collision avoidance constraints.
For simplicity, we consider homogeneous agents under identical constraint sets in this example. 
We collected approximately 2000 samples per scenario with $N_G=200$ and sampling time of $0.1 \; s$, and trained $V_{GT}$ with $h=128, L=2$ using PyTorch. The implementation details for obtaining the GNE solutions and constructing the feature vectors are detailed in Sec. \ref{sec:int_implemnetation}.

Again, to study the impact of the $V_{GT}$, we consider two different terminal cost-to-go functions, 
   \begin{equation} \label{eq:V_int}
    V=
    \begin{cases}
      -V_{GT}(\Tilde{z}^i,\hat{\Tilde{z}}^{\neg i}), \\
      V_{MP}(z^i,\hat{z}^{\neg i})= -(s^i+\hat{s}^{\neg i}),\quad i = 1,2,
    \end{cases}
  \end{equation}
where $V_{MP}$ is a naive maximum progress function that maximizes the progress of agent $i$ without considering the interaction with other agents. The optimization problem \eqref{opt:mpc_pol} is formulated in CasADi and solved using IPOPT. In the cooperative two-vehicle un-signalized intersection example, we assume the vehicles are connected and communicate their route information and planned optimal trajectories, $\hat{\mathbf{z}}^i=\{z^{i,\star}_{k|t}\}^{t+N}_{k=t}$ without delay, allowing $\Tilde{x}^i_{t+N}$ to be computed exactly at time $t$.

\subsubsection{Numerical Results}
\begin{figure*}
    \centering
\includegraphics[width=0.9\textwidth]{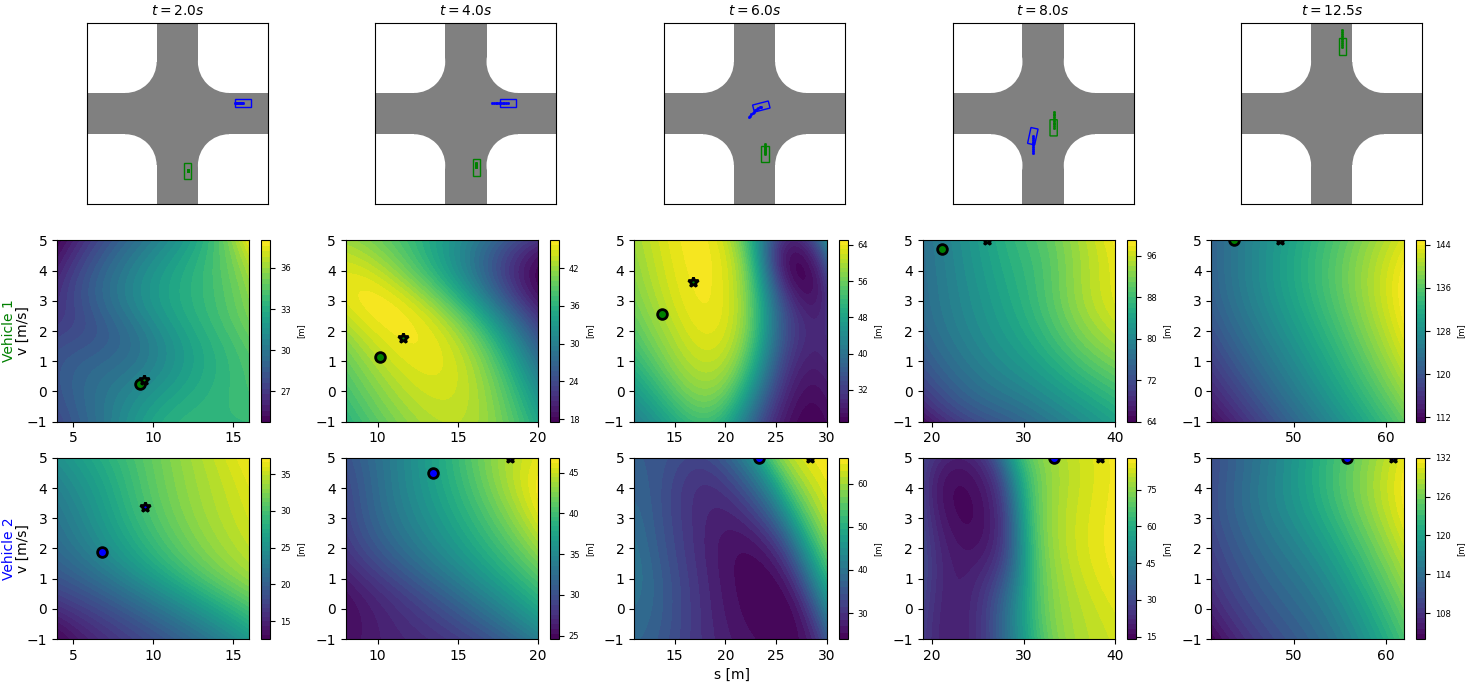}
    \caption{\small{In a simulation experiment, Vehicle 1 (green) on route \textit{SN} and Vehicle 2 (blue) on route \textit{ES} navigate through an intersection using $V_{GT}$ (Top). Colored squares represent each vehicle's planned trajectories. The middle and bottom rows display the contour of $V_{GT}$ as perceived by Vehicle 1 and 2, respectively. Colored circles indicate current states at each time instance, and colored stars denote the planned terminal state at time $t+N$.}}
    \label{fig:int_GT_example}
\end{figure*}

We highlight the differences in how $V_{GT}$ and $V_{MP}$ guide the short-horizon MPC solutions through an intersection from the same initial condition. In Fig. \ref{fig:int_GT_example}, we show Vehicles 1 (green) on route \textit{SN} and Vehicle 2 (blue) on \textit{ES} navigating through an un-signalized intersection starting from rest, where both vehicles are controlled using identical MPC policies \eqref{opt:mpc_pol} with the same $V_{GT}$. However, the input features to $V_{GT}$ are constructed from each vehicle's perspective by concatenating their respective features, as specified in \eqref{eq:int_feature}. 
Although both vehicles share the same policy and value function, their unique perspectives lead to different inputs and outputs. This enables each to follow the collaborative strategy while exhibiting distinct behaviors tailored to their individual contexts.
For instance, at \( t = 4\;\mathrm{s} \), as both vehicles approach the intersection, the level curves of \( V_{GT} \) for Vehicle 1 show a preference for the terminal states in \( s^{1}_{t+N} \in [10,\,13]\;\mathrm{m} \) and \( v^{1}_{t+N} \in [2,\,3]\;\mathrm{m/s} \), whereas Vehicle 2's level curves indicate a preference for terminal states in \( s^{2}_{t+N} \in [17,\,20]\;\mathrm{m} \) and \( v^{2}_{t+N} \in [4,\,5]\;\mathrm{m/s} \). This illustrates how \( V_{GT} \) is guiding the MPC planner for Vehicle 1 to accelerate slowly and yield, while \( V_{GT} \) is guiding the Vehicle 2 to accelerate quickly and pass the intersection first. At \( t = 6\;\mathrm{s} \), when Vehicle 2 reaches the middle of the intersection at its maximum speed of \( 5\;\mathrm{m/s} \), the level curves of \( V_{GT} \) for Vehicle 1 show a preference for higher speed and progress, guiding Vehicle 1 to start accelerating. After Vehicle 2 completes its left turn and is no longer in the path of Vehicle 1, $V_{GT}$ guides Vehicle 1 to accelerate to maximum speed, allowing both vehicles to pass the intersection by \( t = 12.5\;\mathrm{s} \).

\begin{figure*}
    \centering
\includegraphics[width=0.9\textwidth]{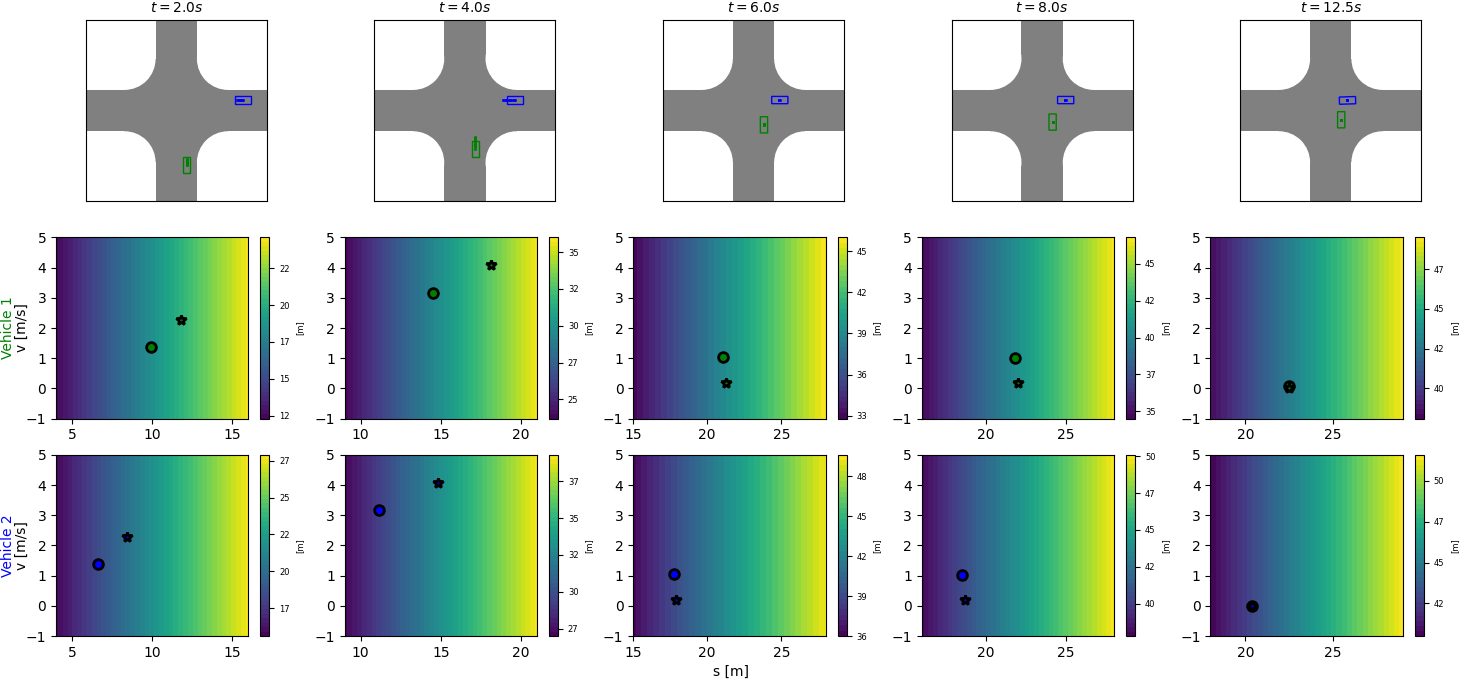}
    \caption{\small{In a simulation experiment, Vehicle 1 (green) on route \textit{SN} and Vehicle 2 (blue) on route \textit{ES} navigate through an intersection using $V_{MP}$ (Top), reaching a gridlock. Colored squares represent each vehicle's planned trajectories. The middle and bottom rows display the contour of $V_{MP}$ as perceived by Vehicle 1 and 2, respectively. Colored circles indicate current states at each time instance, and colored stars denote the planned terminal state at time $t+N$. The initial conditions are identical to that of in Fig. \ref{fig:int_GT_example}}}
    \label{fig:int_MP_example}
\end{figure*}
In Fig. \ref{fig:int_MP_example}, when both MPC planners are guided by $V_{MP}$, they prioritize maximizing progress without accounting for their actions' impact on other agents, as shown by the linear level sets in the middle and bottom rows. This leads to planner infeasibility due to unavoidable collisions from inertia, triggering the safety controller to apply maximum braking at $t=6\; \mathrm{s}$. As a result, the vehicles become stationary and gridlocked, unable to proceed due to coupled collision avoidance constraints.
While extending the prediction horizon could mitigate this issue, it increases computational complexity. Even so, the naive \textit{predict-then-plan} approach can still encounter infeasibility and gridlock in this intersection scenario, as demonstrated in the \href{https://youtu.be/9jlz95Nor2I}{video} \citep{youtube_vid}. 
Furthermore, simulation experiment results demonstrating the effectiveness of $V_{GT}$ in emulating coordinated behaviors in different scenarios are reported in Sec. \ref{sec:additional_results}.

For each scenario $m\in\mathbb{I}_1^8$ as defined in Table \ref{tab:scenario_table} in Sec. \ref{sec:int_implemnetation}, route combinations are sampled from each $\mathcal{S}_m$ to obtain the corresponding precomputed reference trajectory, $z_r$, for the route. Initial $s^1_0, s^1_0 \sim U(0,s_{int}/2)$ are sampled, and $v^1_0,v^2_0=0 \;m/s$ is set to ensure vehicle interaction, where $s_{int}$ denotes the longitudinal displacement at the intersection. Closed-loop simulations are then evaluated with both vehicles using identical terminal cost-to-go functions in \eqref{opt:mpc_pol} over a task horizon of $T=150$. The value functions, $V_{MP}$ and $V_{GT}$, are evaluated over 100 initial conditions for each scenario. Performance metrics include \% of feasibility, \% of gridlock (indicating instances where both vehicles are in a stalemate and cannot proceed their routes) and the average solve time of $\eqref{opt:mpc_pol}$.
The feasibility percentage measures the proportion of simulation runs where the MPC planners remain feasible throughout the simulation run. A high feasibility percentage signifies that the terminal cost-to-go function effectively guided the vehicles to navigate collaboratively, avoiding the need for safety controller intervention. Conversely, a lack of clear passing order or collaborative navigation strategy can cause vehicles to enter the intersection simultaneously, resulting in gridlock.
The numerical simulation results in Table \ref{tab:intersection} highlight the effectiveness of the proposed approach. Guided by $V_{GT}$, MPC planners achieve at least 97\% feasibility across all scenarios, consistently outperforming $V_{MP}$. Notably, in scenarios 6 and 7, where the naive maximum-progress terminal cost $V_{MP}$ fails completely, $V_{GT}$ achieves 99\% feasibility with no gridlocks. Vehicles guided by $V_{GT}$ avoid gridlock by adhering to the passing order encoded within the value function, enabling smooth collaborative navigation. In contrast, $V_{MP}$ often results in gridlocks due to its lack of interaction modeling. While $V_{GT}$ introduces higher computation time, alternative NLP solvers could improve efficiency.
\begin{table}[ht]
    \centering
    \caption{\small{Metrics for Two-Vehicle Intersection Navigation Monte Carlo Simulations}} 
    \label{tab:intersection}
    \small{
    \setlength{\tabcolsep}{4pt} 
    \renewcommand{\arraystretch}{0.9} 
    \begin{tabular}{|c|c|c|c|c|c|c|}
        \hline
        \multirow{2}{*}{Scenario} & \multicolumn{2}{c|}{\textbf{Feasibility} (\%)} & \multicolumn{2}{c|}{\textbf{Gridlock} (\%)} & \multicolumn{2}{c|}{\textbf{Avg. Solve Time} (s)} \\ \cline{2-7}
                                   & $V_{MP}$ & $V_{GT}$ & $V_{MP}$ & $V_{GT}$ & $V_{MP}$ & $V_{GT}$ \\ \hline
        1 & 25 & \textbf{99} & 6 & \textbf{0} & \textbf{0.045} & 0.061 \\ 
        2 & 1 & \textbf{97} & 80 & \textbf{0} & \textbf{0.055} & 0.089 \\ 
        3 & 37 & \textbf{100} & 22 & \textbf{0} & \textbf{0.073} & 0.10 \\ 
        4 & 95 & \textbf{100} & \textbf{0} & \textbf{0} & \textbf{0.087} & 0.091 \\ 
        5 & 93 & \textbf{99} & 1 & \textbf{0} & \textbf{0.075} & 0.078 \\ 
        6 & 0 & \textbf{99} & 43 & \textbf{0} & \textbf{0.082} & 0.10 \\ 
        7 & 0 & \textbf{99} & 100 & \textbf{0} & \textbf{0.077} & 0.098 \\ 
        8 & 1 & \textbf{98} & 57 & \textbf{0} & \textbf{0.083} & 0.084 \\ 
        \hline
    \end{tabular}}
\vspace{-0.5 cm}
\end{table}
\section{Conclusion}
\label{sec:conclusion}
In this work, we presented IGT-MPC, an algorithm that enables decentralized, game-theoretic multi-agent motion planning in competitive and cooperative settings without requiring explicit game-theoretic modeling, which can be computationally expensive.
We demonstrated IGT-MPC through two multi-agent examples: a head-to-head race and a two-vehicle navigation through an un-signalized intersection. The numerical results highlight the effectiveness of the learned value function in guiding MPC to replicate game-theoretic interactions, achieving competitive or coordinated behaviors. Scaling to more than two agents and various environment configurations requires a more extensive dataset that captures rich, game-theoretic interactions, which becomes increasingly computationally expensive to generate. As interactions grow more complex with additional agents, a more sophisticated neural network architecture is needed to capture data patterns. However, this added complexity increases the computational cost of gradient evaluations, leading to longer solution times for the optimization problem. Balancing model complexity with computational efficiency remains a key challenge, which we aim to address in future work.

\bibliography{references.bib}
\section{Appendix}
\subsection{Implementation Details for Intersection Navigation Example}\label{sec:int_implemnetation}
\subsubsection{Cooperative Games} 

 Formulating cooperative games as potential games, if possible, is advantageous over other cooperative game formulations because it captures the effects of all agents' strategy changes within a single global function called a ``potential" function, \textbf{$\Phi(\mathbf{z}, \mathbf{u})$}, where its minimizer corresponds to the GNE \citep{potential_games}. This simplifies the analysis and solving for the GNE as it can be formulated as a single optimization problem. Replacing $\{J^i\}_{i=1}^M$ with $\Phi$The potential function for coordinated navigation at an intersection used in our work is 
\begin{align}\label{eq:potential_func}
     \Phi(\mathbf{z},\mathbf{u})=\sum_{i=1}^2 \sum_{k=0}^{N_{G}-1} \big (\ell^i_k(z^i_{k},u^i_{k}) \big ) -s^i_{t+N_G},
\end{align}
It can be trivially shown that \eqref{eq:potential_func} is a convex, potential function using its definition in \cite{potential_games}. 

\subsubsection{Scenario Encoding}
\begin{table}[!ht]
    \centering
    \caption{\small{Scenarios for Two-vehicle Interactions at an Intersection}}
    \label{tab:scenario_table}    
    \begin{tabular}[c]{| c | c |}
        \hline
        Scenario & Set of Route Combination Tuples ($+, -$)\\
        \hline
        \hline
       $\mathcal{S}_1$  & $\{ (\textit{WE},\textit{NE}),(\textit{NS},\textit{ES}),(\textit{EW}, \textit{SW}),(\textit{SN},\textit{WN})\}$ \\
        \hline
       $\mathcal{S}_2$  & $\{ (\textit{WN},\textit{SW}),(\textit{NE},\textit{WN}),(\textit{ES},\textit{NE}),(\textit{SW},\textit{ES})\}$\\
        \hline
       $\mathcal{S}_3$  & $\{(\textit{WE},\textit{NS}),(\textit{NS},\textit{EW}), (\textit{EW},\textit{SN}),(\textit{SN},\textit{WE})\}$\\
        \hline
    $\mathcal{S}_4$  & $\{(\textit{WN},\textit{EN}),(\textit{NE},\textit{SE}),(\textit{ES},\textit{WS}),(\textit{SW},\textit{NW})\}$ \\
    \hline
    $\mathcal{S}_5$  & $\{ (\textit{WE},\textit{SE}), (\textit{NS},\textit{WS}),(\textit{EW},\textit{NW}),(\textit{SN},\textit{EN})\}$ \\
    \hline
    $\mathcal{S}_6$  & $\{(\textit{WE},\textit{SW}),(\textit{NS},\textit{WN}),(\textit{EW},\textit{NE}),(\textit{SN},\textit{ES})\}$  \\
    \hline
    $\mathcal{S}_7$  & $\{(\textit{WN},\textit{ES}),(\textit{NE},\textit{SW}),(\textit{ES},\textit{WN}),(\textit{SW},\textit{NE})\}$  \\
    \hline
    $\mathcal{S}_8$  & $\{(\textit{WN},\textit{EW}),(\textit{NE},\textit{SN}),(\textit{ES},\textit{WE}),(\textit{SW},\textit{NS})\}$  \\
    \hline
    \end{tabular}
\end{table}

We assume a predetermined reference trajectory is provided based on the desired route, and vehicles must adhere closely to it in compliance with traffic rules. A route is defined by a starting node and a destination node. In a four-way intersection, the nodes are $(\textit{N,W,S,E})$ as depicted in Fig. \ref{fig:scenarios}b. For example, a left turn maneuver from \textit{W} to $N$ node is denoted as route \textit{WN}. Note that no two vehicles can share identical routes. In the intersection example, we employ scenario encoding, where each agent $i$ is assigned a scenario $sc_i$ represented as a signed categorical variable. The magnitude of $sc_i$ denotes a distinct interaction type between the two vehicles, while the sign indicates the vehicle's role in that interaction. For instance, consider the green vehicle (vehicle 1) with a route \textit{SN} and blue vehicle (vehicle 2) with a route \textit{ES} depicted in Fig. \ref{fig:scenarios}b. In this case, the route combination tuple is $(\textit{SN},\textit{ES})\in \mathcal{S}_6$, where $\mathcal{S}_m$ are set of route combination tuples reported in Table \ref{tab:scenario_table} for $m\in\mathbb{I}_1^8$. The tuples in Table \ref{tab:scenario_table} are conventionally ordered where the vehicle of the first element is assigned a positive sign. Therefore, vehicle 1 is assigned $sc_1=+6$ and vehicle 2 is assigned $sc_2=-6$. The magnitude informs us that a vehicle going straight and a vehicle turning left into the same destination node as the vehicle going straight are interacting. Further, the vehicle going straight is given a positive sign while the vehicle turning left is assigned a negative sign to distinguish the role of each vehicle in this unique interaction. Note that scenarios such as both vehicles turning right at the intersection are not considered as vehicles do not interact. 
Also, each scenario exhibits rotational invariance due to rotational symmetry in the symmetric four-way intersection meaning $(\textit{WE},\textit{NE})$ and $(\textit{NS},\textit{ES})$ are identical two-vehicle interactions but rotated $90$ degrees. 

\subsubsection{Training Value Function}
In this example, we train separate $V_{GT}$ for each scenario $\forall m\in \mathbb{I}_1^8$ with distinct datasets only containing samples with corresponding scenario encoding. In the current implementation, we are constrained to lightweight MLP architectures for fast gradient computations necessary for real-time planning (i.e. $\geq 10 \;Hz$) and have trained separate $V_{GT}$ models for each scenario. In future work, we aim to accelerate the solution of \eqref{opt:mpc_pol} by using faster NLP solvers to enable the use of more complex architectures for representing $V_{GT}$, which would improve generalization across multiple scenarios and intersection geometries, thereby enhancing the scalability of our method.

\subsection{Additional Simulation Results} \label{sec:additional_results}

\begin{figure*}
    \centering
\includegraphics[width=0.89\textwidth]{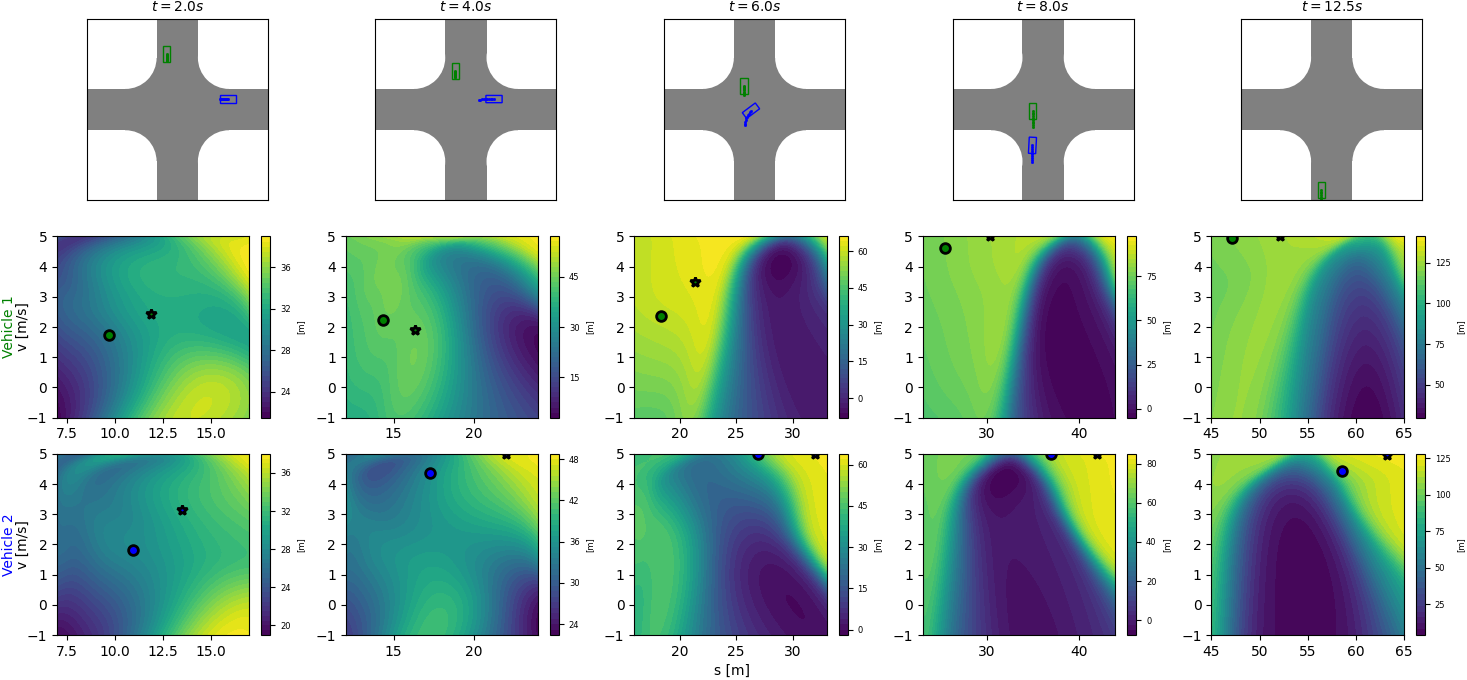}
\caption{\small{In a simulation experiment, Vehicle 1 (green) on route \textit{NS} and Vehicle 2 (blue) on route \textit{ES} navigate through an intersection using $V_{GT}$ (Top). Colored squares represent each vehicle's planned trajectories. The middle and bottom rows display the contour of $V_{GT}$ as perceived by Vehicle 1 and 2, respectively. Colored circles indicate current states at each time instance, and colored stars denote the planned terminal state at time $t+N$.}}
    \label{fig:scenario1_GT}
\end{figure*}

\begin{figure*}
    \centering
\vspace{-0.5 cm}
\includegraphics[width=0.89\textwidth]{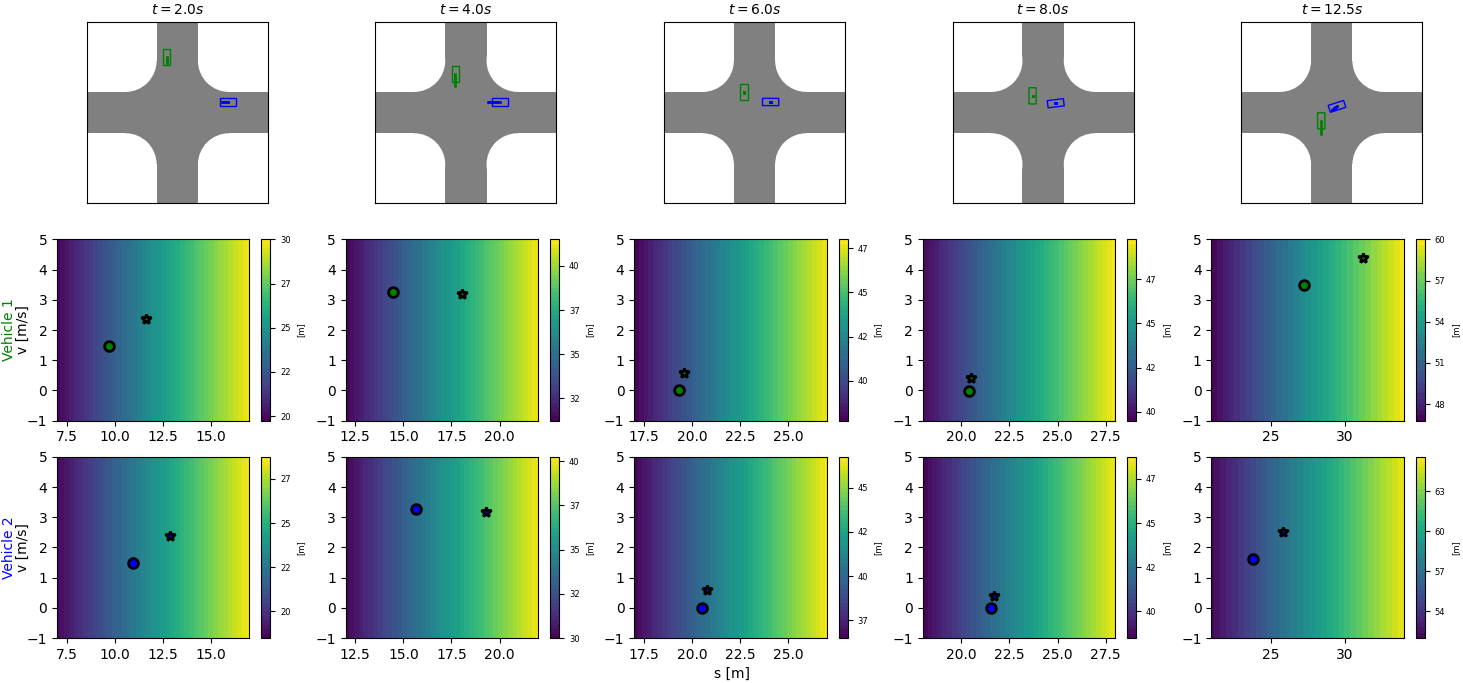}
    \caption{\small{In a simulation experiment, Vehicle 1 (green) on route \textit{NS} and Vehicle 2 (blue) on route \textit{ES} navigate through an intersection using $V_{MP}$ (Top), causing infeasibility of the MPC policy and requiring the safety controller to intervene. Colored squares represent each vehicle's planned trajectories. The middle and bottom rows display the contour of $V_{MP}$ as perceived by Vehicle 1 and 2, respectively. Colored circles indicate current states at each time instance, and colored stars denote the planned terminal state at time $t+N$. The initial conditions are identical to that of in Fig. \ref{fig:scenario1_GT}}}
    \label{fig:scenario1_MP}
    \vspace{-0.5 cm}
\end{figure*}

We further highlight the differences in how $V_{GT}$ and $V_{MP}$ guide the short-horizon MPC solutions through an intersection from the same initial condition in different scenarios. In Fig. \ref{fig:scenario1_GT}, we show Vehicles 1 (green) on route \textit{NS} and Vehicle 2 (blue) on \textit{ES} navigating through an un-signalized intersection starting from rest, where both vehicles are controlled using identical MPC policies \eqref{opt:mpc_pol} with the same $V_{GT}$. As both vehicles approach the intersection at $t=4\;s$, the velocity of vehicle 1 and 2 are $2.3$ and $4.6\;m/s$, illustrating that the $V_{GT}$ is guiding Vehicle 1 to approach the intersection at a slower velocity to yield to Vehicle 2, allowing both vehicles to pass the intersection by $t=12.5\;s$. The blue regions shown in Fig. \ref{fig:scenario1_GT} represent undesirable terminal states that lead to collision or gridlocks, which the MPC policies avoid. In contrast, Vehicle 1 and Vehicle 2 are both approaching the intersection at $3.5\;m/s$ at $t=4\;s$ when guided by a naive $V_{MP}$ in Fig. \ref{fig:scenario1_MP}. The short-sightedness of the MPC policy and unaware of the interaction with other agents, the safety controller intervenes to avoid collision when guided by $V_{MP}$ at $t=6\;s$. 
\begin{figure}
    \centering
\includegraphics[width=0.9\textwidth]{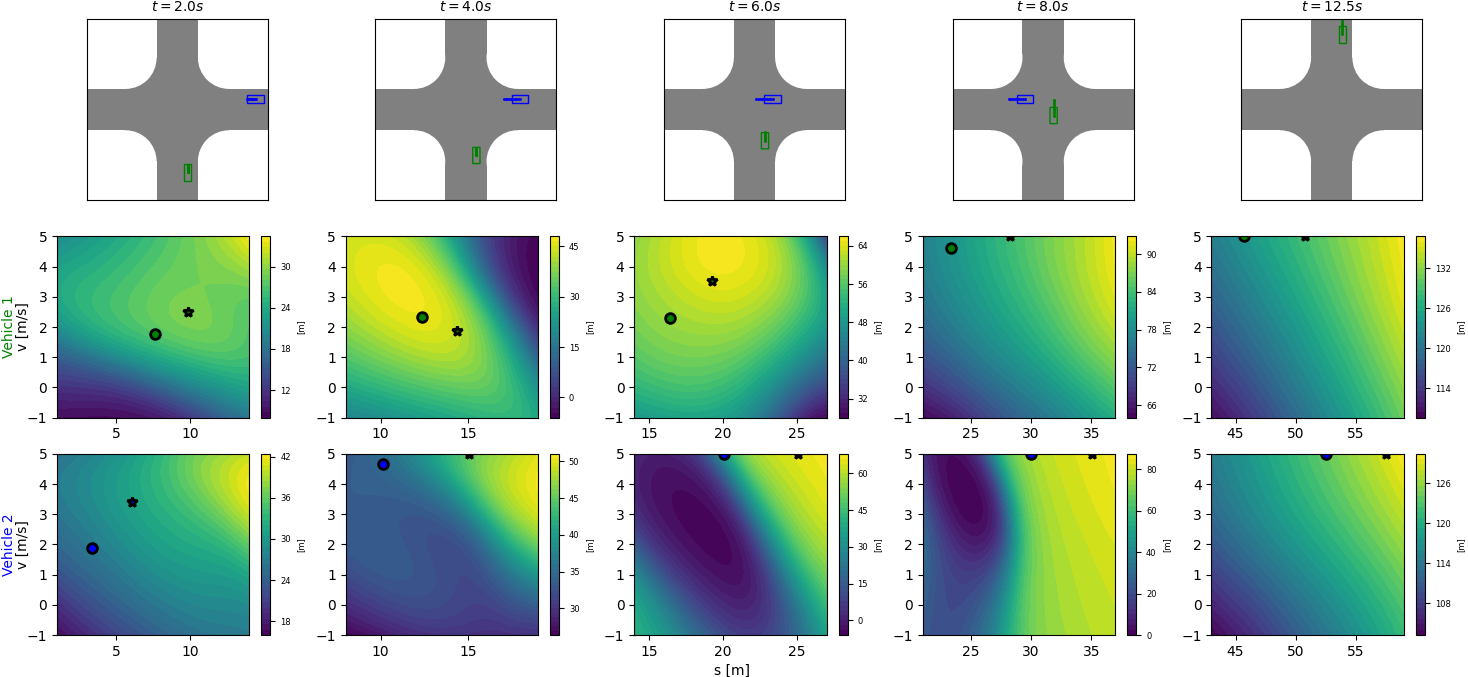}
\caption{\small{In a simulation experiment, Vehicle 1 (green) on route \textit{SN} and Vehicle 2 (blue) on route \textit{EW} navigate through an intersection using $V_{GT}$ (Top). Colored squares represent each vehicle's planned trajectories. The middle and bottom rows display the contour of $V_{GT}$ as perceived by Vehicle 1 and 2, respectively. Colored circles indicate current states at each time instance, and colored stars denote the planned terminal state at time $t+N$.}}
    \label{fig:scenario3_GT}
\end{figure}
\begin{figure}
    \centering
\includegraphics[width=0.9\textwidth]{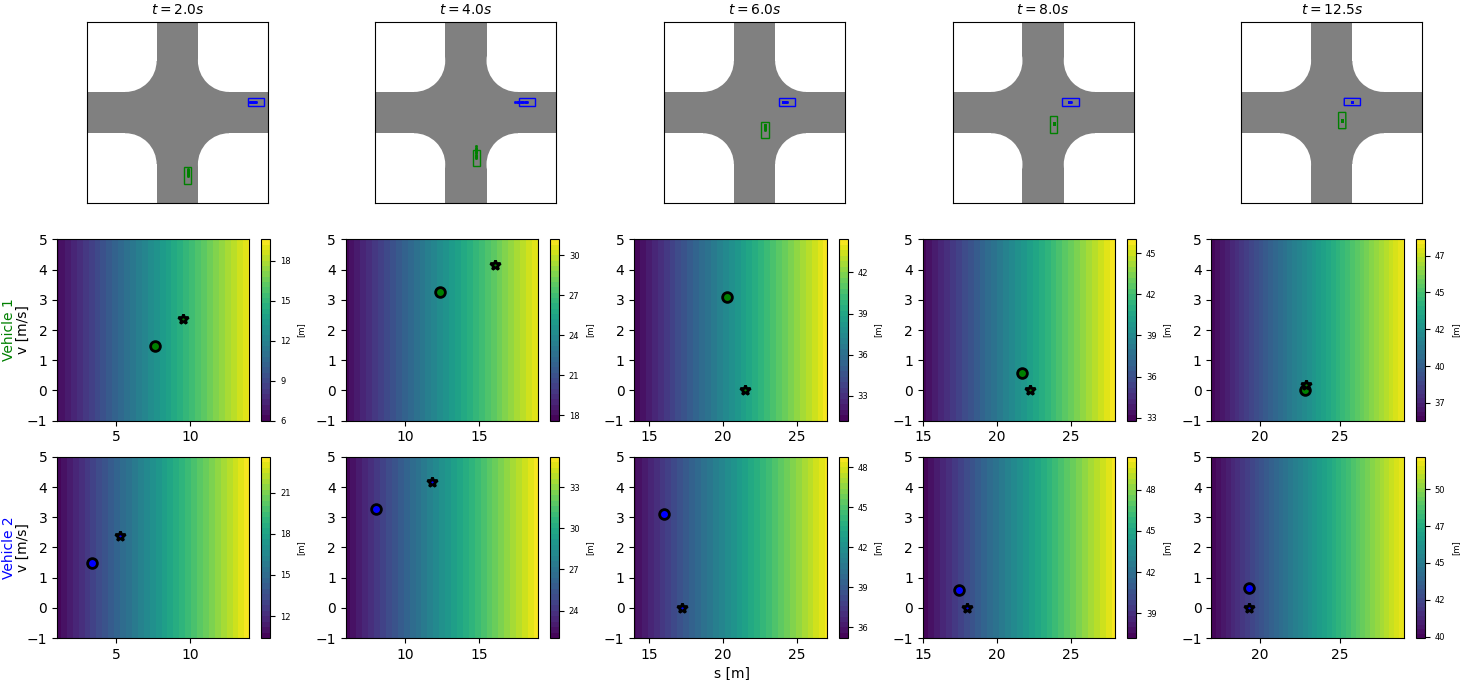}
    \caption{\small{In a simulation experiment, Vehicle 1 (green) on route \textit{SN} and Vehicle 2 (blue) on route \textit{EW} navigate through an intersection using $V_{MP}$ (Top), reaching a gridlock. Colored squares represent each vehicle's planned trajectories. The middle and bottom rows display the contour of $V_{MP}$ as perceived by Vehicle 1 and 2, respectively. Colored circles indicate current states at each time instance, and colored stars denote the planned terminal state at time $t+N$. The initial conditions are identical to that of in Fig. \ref{fig:scenario3_GT}}}
    \label{fig:scenario3_MP}
\end{figure}
Lastly, we show Vehicles 1 (green) on route \textit{SN} and Vehicle 2 (blue) on \textit{EW} navigating through an un-signalized intersection starting from rest, where both vehicles are controlled using identical MPC policies \eqref{opt:mpc_pol} with the same $V_{GT}$ in Fig. \ref{fig:scenario3_GT}. Interestingly, the level curves for Vehicle 1's $V_{GT}$ at $t=4\;s$ show a preference for slowing down within its reachable set in the $s-v$ state space while accelerating (high $s,v$ region) reflect undesirable states. At $t=6\;s$, Vehicle 1 $V_{GT}$'s topology changes and shows a preference for accelerating only after Vehicle 2 is passing the intersection.

In contrast, in Fig. \ref{fig:scenario3_MP} when both MPC planners are guided by $V_{MP}$, it prefers to maximize its progress without considering the impact of their actions on other agents as illustrated by linear level sets in the middle and bottom row of Fig. \ref{fig:scenario3_MP}. Consequently, the vehicles become stationary and reach a gridlock.
\end{document}